\documentclass[sigconf,nonacm]{acmart}
\usepackage{multirow}
\usepackage{makecell} 

\begin{document}

\title{Exploring Finetuned Audio-LLM on Heart Murmur Features}

\author{Adrian Florea}
\email{anf2143@columbia.edu}
\orcid{1234-5678-9012}
\affiliation{%
  \institution{Columbia University}
  \city{New York}
  \state{New York}
  \country{USA}
}

\author{Xilin Jiang}
\email{xj2289@columbia.edu}
\affiliation{%
  \institution{Columbia University}
  \city{New York}
  \state{New York}
  \country{USA}
  }

\author{Nima Mesgarani}
\email{nima@ee.columbia.edu}
\affiliation{%
  \institution{Columbia University}
  \city{New York}
  \state{New York}
  \country{USA}
  }

\author{Xiaofan Jiang}
\email{jiang@ee.columbia.edu}
\affiliation{%
  \institution{Columbia University}
  \city{New York}
  \state{New York}
  \country{USA}
  }

\renewcommand{\shortauthors}{Florea et al.}

\begin{abstract}
Large language models (LLMs) for audio have excelled in recognizing and analyzing human speech, music, and environmental sounds. However, their potential for understanding other types of sounds, particularly biomedical sounds, remains largely underexplored despite significant scientific interest.  In this study, we focus on diagnosing cardiovascular diseases using phonocardiograms, i.e., heart sounds. Most existing deep neural network (DNN) paradigms are restricted to heart murmur classification (healthy vs unhealthy) and do not predict other acoustic features of the murmur such as timing, grading, harshness, pitch, and quality, which are important in helping physicians diagnose the underlying heart conditions. We propose to finetune an audio LLM, Qwen2-Audio, on the PhysioNet CirCor DigiScope phonocardiogram (PCG) dataset and evaluate its performance in classifying 11 expert-labeled murmur features. Additionally, we aim to achieve more noise-robust and generalizable system by exploring a preprocessing segmentation algorithm using an audio representation model, SSAMBA. Our results indicate that the LLM-based model outperforms state-of-the-art methods in 8 of the 11 features and performs comparably in the remaining 3. Moreover, the LLM successfully classifies long-tail murmur features with limited training data, a task that all previous methods have failed to classify. These findings underscore the potential of audio LLMs as assistants to human cardiologists in enhancing heart disease diagnosis.
\end{abstract}

\maketitle

\section{Introduction}
Cardiovascular diseases are the leading cause of death worldwide, claiming a life every 33 seconds in the USA \cite{cdc_wonder_2024}. Auscultation, performed with a stethoscope, is a key practice for examining the circulatory system, allowing physicians to detect abnormal heart and breathing functions by listening to blood flow and heart valve activity. A thorough cardiovascular exam is essential for identifying heart murmurs and associated diseases \cite{pelech2004physiology, Power2020MOLE}. Murmurs are important diagnostic observations, since many conditions present with characteristic murmurs. Distinctive features of phonocardiogram (PCG) can be very helpful in narrowing a differential diagnosis. Physicians describe murmurs by evaluating their timing in the cardiac cycle, intensity, location, duration, configuration, pitch, and quality \cite{Power2020MOLE, pelech2004physiology}. The careful auscultation allows a very close diagnosis, which is the basis for better care.


Recent advancements in large language models (LLMs) for healthcare \cite{Luo_2022, li2023chatdoctormedicalchatmodel, singhal2023expertlevelmedicalquestionanswering, wang2023clinicalgptlargelanguagemodels} enhance tasks like medical QA, record analysis, and licensing exam preparation. While some models \cite{zhang2023huatuogpttaminglanguagemodel, wang2023clinicalgptlargelanguagemodels, chen2024huatuogptii} provide diagnostic insights directly to patients, they risk harm without medical oversight. Conversely, others works support the use of LLMs as supportive tools to physicians who can validate and safely implement the conclusions these models generate \cite{xie2024llms, chen2024codinterpretablemedicalagent}. In the medical domain, multimodal LLMs have been studied in time-series contexts such as electrocardiograms (ECG) \cite{li2023frozenlanguagemodelhelps}, MRI \cite{kanzawa2024automated}, and radiology \cite{ray2024integrating}. These assistive models support clinicians by synthesizing large volumes of information, generating detailed medical summaries, and aiding in clinical decision-making. 

\begin{figure*}[t]
  \centering
  \includegraphics[width=1.0\textwidth]{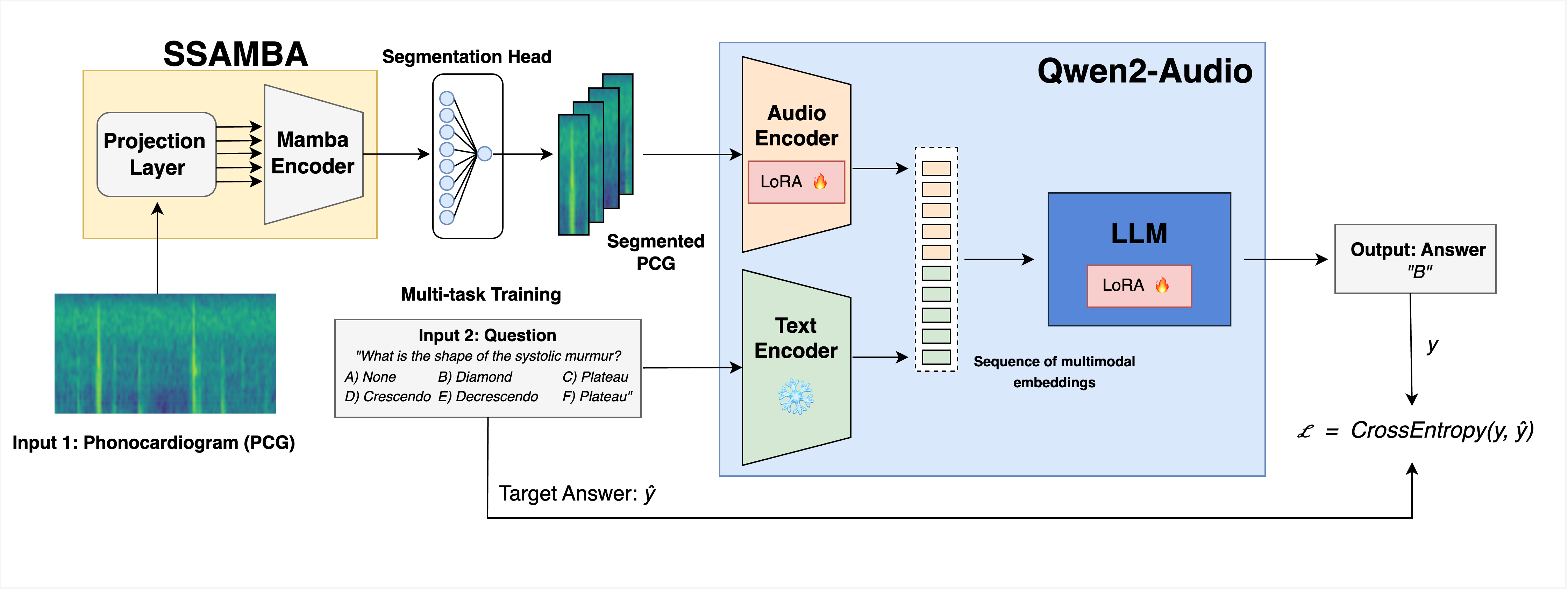} 
  \caption{Proposed PCG system incorporating a finetuned audio LLM Qwen2-Audio model with a front-end segmentation network}
  \label{framework}
\end{figure*}

Moreover, significant advances have been made in the development of audio LLMs \cite{chu2023qwenaudioadvancinguniversalaudio, kong2024audioflamingonovelaudio, tang2024salmonngenerichearingabilities}, demonstrating remarkable achievements in understanding diverse speech signals, complex acoustic and semantic reasoning, and performing speech signal analysis. However, to the best of our knowledge, audio LLMs have not yet been adapted for acoustic health tasks such as PCG analysis \cite{jin2023largemodelstimeseries}. This is the first work that systemically assesses the potential of a state-of-the-art audio LLM as a physician tool for analyzing PCGs. We first leverage the unique repeatative and segmentable nature of PCGs and use a state-space audio representation model, SSAMBA \cite{shams2024ssambaselfsupervisedaudiorepresentation}, to segment the PCGs. We then employ Qwen2-Audio \cite{qwen2audio} for PCG encoding and medical reasoning. Our key contribution is finetuning and evaluating the audio LLM to predict a comprehensive set of 11 physiological PCG features, achieving performance that matches or surpasses traditional methods across all of them. Additionally, the inclusion of a segmentation model enhances robustness, enabling our approach to perform effectively on unseen datasets where PCGs are collected using varied methods.




\section{Background}
\subsection{Heart Sounds}
A normal cardiac cycle includes S1, the first heart sound, caused by the closure of the mitral and tricuspid valves at systole's start, and S2, the second heart sound, caused by the closure of the aortic and pulmonary valves between systole and diastole \cite{bourouhou2020heart}. Additionally, extra heart sounds, i. e., the third heart sound S3 and the fourth heart sound S4, can occur in both normal and pathological conditions \cite{varghees2017effective}. Murmurs, caused by turbulent blood flow in the heart system, are identified as abnormal sounds, and are crucial for diagnosing cardiovascular diseases. Clinically, murmurs consist of two types: systolic murmurs and diastolic murmurs. Aortic stenosis, mitral regurgitation, and tricuspid regurgitation occur during systole, while mitral stenosis and tricuspid stenosis occur during diastole \cite{noor2013heart}.

\subsection{Related Works}
Classifying PCG signals using discriminative machine learning (ML) models has become standard practice, relying on features like time-domain characteristics (e.g., envelope, energy, amplitude) and spectral attributes to identify pathological conditions \cite{ren2024comprehensive, varghees2017effective, langley2017heart}. Other approaches delve into deeper representations, including graph-based features \cite{tuncer2021application}, autoencoder-derived features \cite{humayun2018ensemble}, and sparse coefficients \cite{whitaker2017combining}. Recently, end-to-end ML methods, such as 1D CNNs, TCNs, and GRU-RNNs, have emerged, operating directly on raw audio or mel spectrograms to eliminate the need for manual feature engineering \cite{xiao2020heart, hu2021automatic}. These models show significant potential in detecting subtle cardiac anomalies, such as early indicators of heart failure \cite{gao2020gated}, by learning relevant patterns directly from data. The complexity of feature selection and end-to-end architectures  demands robust frameworks that can truly make use of such diversified signal properties. Furthermore, ML models aim to achieve more noise-robust results by employing a segmentation preprocessing step that splits an entire PCG into smaller segments (S1,systole, S2, and diastole) \cite{ren2024comprehensive}. Studies have show that segmentation of S1 and S2 phases improve classification performance \cite{dissanayake2020robust} and reduce the need for labeled data \cite{segmentation_dl}.


\begin{table*}[h!]
\centering
\normalsize
\setlength{\tabcolsep}{10pt} 
\caption{Comparison of systolic feature classification accuracy between our model and state-of-the-art methods. NS denotes no segmentation, while W.S indicates the use of a segmentation front-end.}
\label{tab:systolic_accuracy}
\begin{tabular}{lccccc}
\toprule
\multirow{2}{*}{\textbf{Features (\% accuracy)}} & \multicolumn{5}{c}{\textbf{Systolic Phase Feature}}  \\
 & \textbf{Timing} & \textbf{Shape} & \textbf{Grading} & \textbf{Pitch} & \textbf{Quality} \\
\midrule
\textbf{Deep CardioSound \cite{guo2022deep}} & 96.6 & 96.3  & \textbf{96.6} & 96.6  & 96.5 \\
\textbf{This Work N.S}  & \textbf{100} & 99.7 & 33.4 & 99.7 & 99.2\\
\textbf{This Work W.S}  & \textbf{100} & \textbf{100} & 33.4 & \textbf{100} & \textbf{100}\\
\bottomrule
\end{tabular}
\end{table*}

Another important achievement in this regard is the use of transfer learning to overcome the limitation of small annotated datasets. Transfer learning involves finetuning models trained on a large generic dataset for specific tasks. In some studies \cite{niizumi2024exploring, ren2018learning,alaskar2019implementation,koike2020audio}, pretrained models are primarily learnt on an image dataset (ImageNet) \cite{deng2009imagenet} and an audio dataset (AudioSet) \cite{gemmeke2017audio}. Finetuned models have even outperformed pretrained models as they adapt to the data distribution of heart sound datasets. Transfer learning is highly effective because finetuning pretrained models enables strong generalization to new data. Consequently, transfer learning reduces the data scarcity problem and improves model accuracy, hence is a crucial approach for developing robust PCG classification systems \cite{ren2024comprehensive}. 

While most ML research in heart sound analysis has focused on binary classification of heart murmurs as either healthy or unhealthy, other murmur characteristics, such as timing, grading, shape, quality, and pitch, have received comparatively limited attention; evident by works that focus on these features for systolic murmurs \cite{wu2024heart,guo2022deep}. These characteristics are clinically significant as they provide deeper insights into the underlying pathophysiology of cardiac abnormalities. A notable study introduced models incorporating feature attention modules to tackle these nuanced classifications, achieving state-of-the-art performance \cite{guo2022deep}. Despite these advancements, research exploring these additional features -- especially diastolic features -- remains sparse, underscoring the need for more comprehensive investigations into the diverse attributes of heart murmurs. \vspace{-7pt}

\section{Methods}
\subsection{Models}
Our proposed PCG multi-feature classification system effectively utilizes an Audio LLM finetuned for PCG for feature extraction and medical reasoning, along with a PCG segmentation front-end to separate heartbeats and non-heartbeats. The entire system is shown in Figure \ref{framework}.

\subsubsection{PCG Segmentation Model}
Mamba state space models (SSMs) have demonstrated an exceptional capacity in capturing temporal dependencies across local and global scales \cite{gu2024mambalineartimesequencemodeling}. Building on this, Mamba-based audio encoding models like SSAMBA \cite{shams2024ssambaselfsupervisedaudiorepresentation} further enhance temporal modeling by dividing audio spectrograms into patches and processing them in both forward and backward directions. This modeling approach, combined with the audio representations learned from the AudioSet \cite{gemmeke2017audio}, can be effectively adapted to PCG signals. To achieve this, we added a linear head on top of SSAMBA for segmenting heartbeats into occurrences and silences and finetuned the entire model using only one-third of the training set:

\begin{equation}
L = - \frac{1}{T} \sum_{t=1}^{T} \left[ y_t \log(\hat{y}_t) + (1 - y_t) \log(1 - \hat{y}_t) \right]
\end{equation}
We optimize the binary cross-entropy loss between the predicted probabilities ($\hat{y}_t$) and the ground truth ($y_t$) segmentation labels at each time step in the sequence. This formulation incentivizes the model to assign higher probabilities to time segments corresponding to the onset and offset of each heartbeat ($y_t = 1$) and lower probabilities to other segments ($y_t = 0$).

\subsubsection{Audio LLM Adapted for PCGs}
The model architecture of Qwen2-Audio \cite{qwen2audio} contains an audio encoder based on the Whisper \cite{whisper} speech recognition model and a large language model Qwen-7B \cite{qwen2}, which encompasses a total of 8.2B parameters. Given the paired data \((a, x)\), where \(a\) and \(x\) denote the audio sequences and text sequences respectively, the training objective is to maximize the probability of the next text token. This is expressed as:

\[
P_\theta(x_t \mid x_{<t}, \text{AudioEncoder}_\phi(a)),
\]

conditioning on the audio representations and the preceding text tokens \(x_{<t}\). Here, \(\theta\) represents the trainable parameters of the LLM, and \(\phi\) represents the trainable parameters of the audio encoder. Qwen2-Audio is the state-of-the-art in various speech and audio understanding benchmarks like AIR-Bench, MMAU, and S2TT \cite{qwen2audio}. We finetuned Qwen2-Audio with full low-rank approximation (LoRA) \cite{lora}, which adds low-rank matrics (learnable) to the original weight matrics (frozen) in the audio encoder and the LLM.

\begin{table*}[h!]
\centering
\normalsize
\setlength{\tabcolsep}{10pt} 
\caption{Comparison of diastolic features classification and weighed murmur classification accuracy between our model and state-of-the-art methods}
\label{tab:diastolic_accuracy}
\begin{tabular}{lcccccc}
\toprule
\multirow{2}{*}{\textbf{Features (\% accuracy)}} & \multicolumn{5}{c}{\textbf{Diastolic Phase Feature}} & \textbf{Murmur} \\
 & \textbf{Timing} & \textbf{Shape} & \textbf{Grading} & \textbf{Pitch} & \textbf{Quality} & \textbf{W.acc} \\
\midrule
\textbf{M2D + AST \cite{niizumi2024exploring}} & -- & -- & -- & -- & -- & \textbf{83.2} \\
\textbf{This Work N.S} & \textbf{100.0} & 99.9 & 39.5 &  \textbf{100.0}  & \textbf{100.0} & 63.7 \\
\textbf{This Work W.S}  & \textbf{100.0}  & \textbf{100.0} & \textbf{42.7} & \textbf{100.0} & \textbf{100.0} & 75.6\\
\bottomrule
\end{tabular}
\end{table*}
\subsection{Data}
The PhysioNet CirCor DigiScope dataset is currently the largest collection of pediatric heart sound recordings \cite{reyna2023heart}. It includes 5,282 recordings captured from the four primary auscultation sites across 1,568 patients, amounting to over 312 hours of heart sound data sampled at 4 kHz. The patient ages span from 0.1 to 356.1 months. Recording durations range from 4.8 to 80.4 seconds, averaging 22.9 seconds with a standard deviation of 7.4 seconds. Using a semi-supervised annotation process, experts meticulously annotated the dataset, detailing characteristics such as the timing, shape, pitch, grading, quality, and location (aortic, pulmonic, tricuspid, or mitral), and heart beat segmentation (systolic, diastolic, S1, and S2 timings) of each murmur. The dataset comprises 74\% murmur-absent cases, 19\% murmur-present cases, and 7\% uncertain cases due to background noise or interference.

\subsection{Dataset}
We stratify patients by class (Present/Absent/Unknown), resample to 16,000 kHz, and create a 75/25\% split for training and testing. We selected 11 classification tasks based on the expert-annotated labels provided in the PhysioNet dataset. These tasks encompass a comprehensive set of murmur characteristics, including both systolic and diastolic features. Specifically, they include murmur weighted accuracy \( W.acc \) as defined in \cite{reyna2023heart}, systolic murmur attributes such as timing, shape, grading, pitch, and quality, as well as corresponding diastolic  timing, shape, grading, pitch, and quality features. Each of these tasks reflects clinically significant dimensions of heart sound analysis, as identified by domain experts. In order to account for model overfitting on the question input, each question in training and testing set is selected randomly from three options with slight variations in how the question is phrased (prior to the presentation of the multiple choices).

To optimize model performance, we finetuned our framework using a multiple-choice (MC), single-answer approach. This design allows the model to focus on selecting the most appropriate label from a predefined set of options for each classification task, aligning with the structured nature of the annotated data. In some cases, the classification task contained less possible features than 6, so the MC distractor solutions were padded with additional instances of incorrect answers.

\section{Results}
We evaluate our system's performance across 11 MC tasks on the PhysioNet CirCor DigiScope test split. Table \ref{tab:systolic_accuracy} compares our performance with the systolic phase feature classification reported in \cite{guo2022deep}. The performance of our model without the frontend segmentation (N.S.) step is comparable to its performance with the frontend segmentation step (W.S), while both obtained optimal classification accuracy for timing, shape, pitch and quality. The classification for both systolic and diastolic grading features exhibits poor performance in classifying between the grading labels i/vi, ii/vi, and iii/vi.  This is likely due to the model's limited understanding of Roman numerals, compounded by the fact that the text encoder was frozen during fine-tuning, restricting its ability to adapt to these specific labels. Though, the segmentation preprocessing step does improve diastolic grading performance above chance level.

Table \ref{tab:diastolic_accuracy} compares our performance for weighted murmur classification accuracy to results reported by \cite{niizumi2024exploring}. Diastolic phase features are long-tail murmur characteristics with limited representation in both the training and test datasets. Consequently, these features have not been extensively studied or incorporated into previous murmur models. We observe murmur classification increase with the use of the segmentation frontend, achieving better performance may be possible through extra training with additional \textit{murmur present} examples. 

Table \ref{tab:dataset_comparison} evaluates the zero-shot normal/abnormal PCG classification performance on the PhysioNet/CinC 2016 PCG dataset (training sets \textit{a} through \textit{f}) \cite{liu2016classification} and Pascal dataset (set A and B) \cite{liu2016open}. Our results demonstrate that our system can effectively differentiate between normal and abnormal PCGs when utilizing the segmentation frontend. This demonstrates that our system is versatile and is capable of effectively handling variations in audio quality, recording conditions, and patient demographics. 

Since our system is the first of its kind, there are no existing models capable of interpreting PCG audio and jointly classifying multiple relevant features in the systolic and diastolic phase as we do. By expanding beyond the 11 features in our benchmark, we hope to inspire further advancements in the use of audio LLMs for comprehensive cardiac sound analysis and improve diagnostic capabilities in cardiovascular care.

\begin{table}[h!]
\centering
\normalsize  
\renewcommand{\arraystretch}{2} 
\setlength{\tabcolsep}{.5pt} 
\caption{Zero-shot murmur classification across different benchmark datasets.}
\label{tab:dataset_comparison}
\begin{tabular}{lccc}
\toprule
\textbf{\makecell{Dataset}} & \makecell{\textbf{PhysioNet} \\ \textbf{CirCor 2016 \cite{liu2016classification}}} & \makecell{\textbf{Pascal} \\ \textbf{Set A \cite{liu2016open}}} & \makecell{\textbf{Pascal} \\ \textbf{Set B \cite{liu2016open}}} \\
\makecell{\textbf{Metric}} & \% Accuracy &  \multicolumn{2}{c}{Normal / Abnormal Precision}  \\
\midrule
\makecell{\textbf{This Work N.S}} & 19.63\% & 48.7 / 0.0 \% & 73.7 / 40.0 \% \\
\makecell{\textbf{This Work W.S}} & 66.33\% & 48.3 / 100.0 \% & 75.3 / 62.5 \% \\
\bottomrule
\end{tabular}
\vspace{15pt} 
\end{table}

\section{Conclusion}
We demonstrate the potential of audio LLMs in advancing the analysis of phonocardiograms for cardiovascular disease diagnosis. By finetuning Qwen2-Audio on the PhysioNet CirCor DigiScope dataset, we show that the proposed system outperforms state-of-the-art models in most key murmur features. The incorporation of a segmentation preprocessing step enhanced the model's robustness and generalizability across diverse datasets. Our findings highlight the ability of audio LLMs to capture nuanced cardiac characteristics, offering valuable support for cardiologists in diagnosing heart conditions with greater precision. By expanding the set of tasks our system can handle in our benchmark, we aim to drive further innovation in using audio LLMs for comprehensive cardiac sound analysis, ultimately enhancing diagnostic tools in cardiovascular care. While LLMs can assist in medical data analysis, human cardiologists remain essential for clinical decision-making, as it requires expert judgment and patient interaction.

\bibliographystyle{ACM-Reference-Format}
\bibliography{sample-base}

\end{document}